\pdfoutput=1
\RequirePackage{ifpdf}
\ifpdf 
\documentclass[pdftex]{sigma}
\else
\documentclass{sigma}
\fi

\numberwithin{equation}{section}

\usepackage{mathtools}
\usepackage{mathrsfs}
\usepackage{phoenician}
\usepackage{multirow}
\usepackage{tikz}
\tikzstyle{every picture}=[level distance = 8mm, baseline=-0.5ex]
\tikzstyle{prop}=[shape=circle,minimum size=6mm, draw=black!80, fill=green!30]

\newcommand{\Oo}{{\mathcal{O}}}
\newcommand{\Ww}{{\mathcal{W}}}
\newcommand{\q}{\textphnc{q}}
\newcommand{\s}{\textphnc{s}}

\begin{document}
\allowdisplaybreaks

\newcommand{\arXivNumber}{2011.13822}

\renewcommand{\thefootnote}{}

\renewcommand{\PaperNumber}{075}

\FirstPageHeading

\ShortArticleName{Resurgent Analysis of {W}ard--{S}chwinger--{D}yson Equations}

\ArticleName{Resurgent Analysis of\\ {W}ard--{S}chwinger--{D}yson Equations\footnote{This paper is a~contribution to the Special Issue on Algebraic Structures in Perturbative Quantum Field Theory in honor of Dirk Kreimer for his 60th birthday. The~full collection is available at \href{https://www.emis.de/journals/SIGMA/Kreimer.html}{https://www.emis.de/journals/SIGMA/Kreimer.html}}}

\AuthorNameForHeading{M.P.~Bellon and E.I.~Russo}

\Author{Marc~P.~BELLON and Enrico~I.~RUSSO}
\Address{Sorbonne Universit\'e, CNRS, Laboratoire de Physique Th\'eorique et Hautes Energies,\\ Paris, France}
\Email{\href{mailto:marc.bellon@upmc.fr}{marc.bellon@upmc.fr}, \href{mailto:erusso@lpthe.jussieu.fr}{erusso@lpthe.jussieu.fr}}

\ArticleDates{Received February 10, 2021, in final form July 30, 2021; Published online August 11, 2021}

\Abstract{Building on our recent derivation of the Ward--Schwinger--Dyson equations for the cubic interaction model, we present here the first steps of their resurgent analysis. In our derivation of the WSD equations, we made sure that they had the properties of compatibility with the renormalisation group equations and independence from a regularisation procedure which was known to allow for the comparable studies in the Wess--Zumino model. The~inter\-actions between the transseries terms for the anomalous dimensions of the field and the vertex is at the origin of unexpected features, for which the effect of higher order corrections is not precisely known at this stage: we are only at the beginning of the journey to use resurgent methods to decipher non-perturbative effects in quantum field theory.}


\Classification{81Q40; 81T16; 40G10}

\renewcommand{\thefootnote}{\arabic{footnote}}
\setcounter{footnote}{0}

\section{Introduction}
Quantum field theories do not produce convergent perturbative series, while these perturbative series are often
the only information accessible up to now through an analytic treatment. Converting these series in numbers
and properties of the theory therefore requires some non trivial summation methods, especially in
the large coupling regime. A very useful method goes through the definition of a Borel transform that will give the
solution through a Laplace integral. However, in many cases, the Borel transform has singularities on the real
axis which make the naive Borel--Laplace summation ambiguous: integration on rotated axis lose the reality
properties of~the original series. In such situations, results with much reduced ambiguities can be obtained
through the use of real averages of the different analytic continuations.

It has been known for a long time that a successful
resummation of a divergent series requires the knowledge of its asymptotic properties and recent
works have tried to quantify the gains which can result from a deeper knowledge of the properties of the Borel
transform. We would single out the paper~\cite{CoDu20} that show how the use conformal maps of the Borel plane
allows for the most precise results.
However, such gains are only possible if the precise structures of the singularities of the Borel transform are
known.

In quantum mechanics, singularities of the Borel transform stem from
the presence of non trivial saddle point of the action functional, dubbed instantons, but in renormalisable
quantum field theories, new singularities appear, related to the behaviour of diagrams with a maximal number of
subdivergences which are called renormalons. Much work has been devoted to the finding of classical field
configurations which could explain these singularities, but with limited success even if we must
cite~\cite{ArUn2012,AsMoSuTa2020}. This work will be based on a quite different approach, in~the spirit
of~\cite{BeCl14,BeCl16}, which made use of the tools of resurgence theory and in particular the alien derivatives, in
the study of the solution of a Schwinger--Dyson equation. This work has been prepared by a number of
studies~\cite{Be10,Be10a,BeCl13} which addressed only what was later recognized as the singularities nearest the origin
of the Borel transform. These previous studies were however limited to the supersymmetric Wess--Zumino model,
first solved perturbatively at high order in~\cite{BeSc08}, or a very special case of the $\phi_6^3$ model we
study here, where the vertex gets no radiative corrections at one loop~\cite{Be10a}.

It is therefore very interesting that our recent work~\cite{BeRu20} gives a system of equations for the
determination of the renormalisation group functions of the $\phi_6^3$ model, that we called
Ward--Schwinger--Dyson equations, most suitable for a resurgent analysis. Indeed, this scheme has the properties
which made the analysis in~\cite{BeCl14} possible, the absence of any explicit regularisation parameter and the
invariance of the solution under the renormalisation group. Indeed one may say that our computations tend to
transform a leading log approximation for the propagator and the vertex in a computation of the leading terms for
the high order terms of the perturbative solution, but with the possibility to go beyond these leading behaviours
through the systematic inclusion of corrections.

It is a quite old observation by Giorgio Parisi that the
renormalisation group equations written for the Borel transform of the propagator imply that at a position $\rho$
in the Borel plane, the propagator has a leading correction proportional to $\big(p^2/\mu^2\big)^{b_1\rho}$, with $b_1$ the
leading coefficient of the $\beta$ function~\cite{Pa79}. These power corrections give rise to new divergences
which seemed impossible to renormalize, in particular in the case of the infrared ones. We will instead show
that these divergences, linked to the poles of the Mellin transforms of the graphs, simply enable us to compute the
singularities of the Borel transforms of the anomalous dimensions.

\section{The Borel--Laplace resummation method}

\subsection{General properties}
We do not have the presumption here to introduce the topic of Borel--Laplace resummation techniques; there are excellent introductions in the literature \cite{Bo11} or \cite{Sa14}. We report here though some basic properties that we will need in the further development.

The {formal Borel transform} is defined on formal series as
 \begin{align*}
 \mathcal{B}\colon\ \big(z^{-1} \mathbb{C}\big[\big[z^{-1}\big]\big],\cdot\big) & \longrightarrow (\mathbb{C}[[\xi]],\star), \\
 \tilde{f}(z) = \frac{1}{z}\sum_{n=0}^{+\infty}\frac{c_n}{z^n} & \longrightarrow \hat{f}(\xi) = \sum_{n=0}^{+\infty}\frac{c_n}{n!}\xi^n.
 \end{align*}
 Let $\tilde{f}$, $\tilde{g}$ in $z^{-1} \mathbb{C}\big[\big[z^{-1}\big]\big]$ be two formal series and $\hat{f}$, $\hat{g}$ in $\mathbb{C}[[\xi]]$ be their Borel
 transforms. The following properties hold
 \begin{gather*}
 \mathcal{B}\big(\tilde f.\tilde g\big) = \hat f\star \hat g, \qquad
 \mathcal{B}\big(\partial\tilde f\big) = -\zeta\hat f, \qquad
 \mathcal{B}\big(z^{-1}\tilde f\big) = \int\!\hat f,
 \\
 \tilde{f}(z)\in z^{-2} \mathbb{C}\big[\big[z^{-1}\big]\big] \Longrightarrow \mathcal{B}\big(z\tilde f\big)=\frac{{\rm d}\hat f}{{\rm d}\zeta},
 \end{gather*}
with the derivatives and the integral defined term by term and $\star$ denoting the convolution product of formal series.
If $\hat f$ and $\hat g$ are convergent,
\begin{gather*}
 \hat f\star \hat g(\zeta) = \int_0^\zeta\hat f(\eta) \hat g(\zeta-\eta)\,{\rm d}\eta
\end{gather*}
for $\zeta$ in the intersection of the convergence domains of $\hat f$ and $\hat g$. The analytic continuation of~the convolution product on a given path can also be expressed through such an integral, but the integration path is in general much more complex than the path on which the analytic continuation is taken.

The definition of the Borel transform can be extended to series with constant terms through the introduction of a unit $\delta$ for the convolution product.
The constant function equal to the constant $a$ is mapped to $a \delta$ and
then the whole space of formal series $\mathbb{C}[[z^{-1}]]$ can be mapped by linearity.

A formal series $\tilde f(z) = \frac{1}{z}\sum_{n=0}^{+\infty}\frac{a_n}{z^n}$ is {1-Gevrey} if
\begin{gather*}
 \exists A,B>0\colon\ |a_n|\leq AB^n n!\quad\forall n\in\mathbb{N}.
\end{gather*}
In this case, we write $\tilde f(z)\in z^{-1}\mathbb{C}\big[\big[z^{-1}\big]\big]_1$.
In this case and only in this case, its Borel transform has a finite radius of convergence and we denote by~$\mathbb{C}\{\zeta\}$ the space of such functions.

The Borel transform can be inverted through the Laplace transform.
Let $\theta\in[0,2\pi[$ and set $\Gamma_\theta:=\big\{R{\rm e}^{{\rm i}\theta},R\in[0,+\infty[\big\}$. Let $\hat f\in\mathbb{C}\{\zeta\}$ be a germ admitting an
analytic continuation in an~open subset of $\mathbb{C}$ containing $\Gamma_\theta$ and such that
\begin{gather} \label{eq:exp_bound}
\exists c\in\mathbb{R},\, K>0\colon\ \big|\hat f(\zeta)\big|\leq K {\rm e}^{c|\zeta|}
\end{gather}
for any $\zeta$ in $\Gamma_\theta$. Then the Laplace transform of $\hat f$ in the direction $\theta$ is defined as
\begin{gather*}
\mathcal{L}^{\theta}\big[\hat f\big](z) = \int_0^{{\rm e}^{{\rm i}\theta}\infty}\hat f(\zeta){\rm e}^{-\zeta z}\,{\rm d}\zeta.
\end{gather*}
When the bound~\eqref{eq:exp_bound} is verified, this expression is finite in the half-plane ${\rm Re}\big(z{\rm e}^{{\rm i}\theta}\big)>c$
and therefore defines an analytic function of~$z$ in this domain, which is called a Borel sum of~$\tilde f$.

For a formal series $\tilde f(z)\in z^{-1}\mathbb{C}\big[\big[z^{-1}\big]\big]$ with a non-zero radius of convergence,
equation~\eqref{eq:exp_bound} is true for all $\theta\in[0,2\pi[$ and its Borel sum in any direction coincide
with the usual sum of the series. For more general Borel summable series, many interesting phenomena can arise,
such as the Stokes phenomenon: the singularities of the Borel transform imply differences between the Borel sums
defined in directions separated by these singularities and even in cases, where the condition~\eqref{eq:exp_bound}
is satisfied for all directions, the Borel sum will differ from its analytic continuation in a path around
infinity, giving a non trivial monodromy.
These problems are at the heart of the renewed interest in summability techniques in
particular in the physics community~\cite{AnSc13,AnScVo12,DuUn14,Marino_Reis_2019}.

\subsection{Resurgent functions and alien derivatives} \label{alien}
Borel summation heavily relies on the possibility of analytically continuing the Borel transform on a
neighbourhood of the integration axis. In fact, the situation is usually better, with a~continuation possible
in the whole complex plane minus some set of singularities.
These singularities can be studied through alien derivatives. The alien operator $\Delta_\omega$ extracts
the singularity around~$\omega$ of the Borel transform and translates
it to the origin. Some care must be taken when~$n$ singularities lie on
the segment $[0,\omega]$ since there is no longer a canonical analytic continuation of the Borel transform
to the neighborhood of $\omega$: each singularity can be avoided in two dif\-fe\-rent ways resulting in $2^n$
possible analytic continuations. It is easy to show that in
the simple case without singularities between 0 and $\omega$, $\Delta_\omega$ is a derivation with respect
to the convolution product. With adequate weighting of the different paths to $\omega$, this can be made true
in the general case with any number of singularities on the path.

{\samepage
The alien operators corresponding to the singularities nearest the origin have special sig\-ni\-fi\-cance:
being the one which limit the convergence radius of the Taylor series, they give the dominant behaviour in their
high orders. A singularity at a point $\omega$ gives a contribution with a ratio $1/\omega$ between
successive orders. The first correction is important, since it is related to the kind of singularity. In the
case of power laws singularities $(\xi - \omega)^{-\alpha}$, the exponent $\alpha$ translates into a
$n^{\alpha-1}$ factor in the $n$-th coefficient of the Taylor series.\footnote{If $\alpha$ is a nonpositive
integer, $(\xi-\omega)^{-\alpha}$ is not singular at $\omega$ and the function has to be multiplied by
$\log(\xi-\omega)$ to obtain the $n^{\alpha-1}$ behaviour.} Translating to the original asymptotic series,
this can be seen to mean that we have the following asymptotic ratio of successive terms of the coefficients
\begin{gather} \label{c_asymp}
 \frac {c_{n+1}}{c_n} = \frac 1 \omega \bigl( n + \alpha + \mathcal{O}\big(n^{-1}\big) \bigr).
\end{gather}
Beware that this formula is for the $c_n$'s as defined in the previous subsection, with an index shifted by 1
with respect to other conventions. Many studies have focused on obtaining this type of relation, but the
properties of the alien derivatives allows for a more efficient derivation.
\looseness=1

}

Since the alien derivative involves a translation and the ordinary derivative becomes the multiplication
by~$-\zeta$ for the Borel transform, these two derivatives do not commute: we have
\begin{gather*}
	[ \partial, \Delta_\omega] = \omega \Delta_\omega. 
\end{gather*}
Alien derivatives can be considered to act on the formal power series in $z^{-1}$ and one generally keeps the
same notation, since confusion is not possible. In this case, we can multiply the operator $\Delta_\omega$ by
a transmonomial to obtain $ \dot\Delta_\omega \equiv {\rm e}^{-\omega z} \Delta_\omega $. This modified operator
$\dot\Delta_\omega$ now commutes with the derivative and since it is a derivation, $\dot\Delta_\omega f$
represent a possible deformation of the solution $f$ of a system of differential equations.

When a system of equations is given, alien derivatives can therefore be determined in two stages. One first
determines all possible deformations of a solution involving transmonomial factors ${\rm e}^{-\omega_i z}$, extending
the solution to a transseries. This transseries will depend on para\-me\-ters~$c_i$ and the possible
alien derivatives are given by bridge equations, which express each alien derivative through the action of some
differential operator in the parameters~$c_i$. It is this computation scheme that we will use throughout,
rather than attempting to obtain directly the asymptotic ratios in equation~\eqref{c_asymp}.

\section[WSD equations for phi 3 6]{WSD equations for $\boldsymbol{\phi^3_6}$}

We introduced the Ward--Schwinger--Dyson scheme (often abbreviated to WSD in the following) in the case of the
model $\phi^3_6$ in~\cite{BeRu20}. We here recall only the equations which will be studied. For~their origin and
possible extensions, the reader is invited to go back to this previous work.
They can be considered as variations on the Schwinger--Dyson equations
written in terms of derivatives of the effective action~\cite{DeWitt65}.
The lowest order primitive terms of the
Ward--Schwinger--Dyson equations for $\phi^3_6$ have the following diagrammatic form
\begin{equation}\label{WSDdiag}
 \begin{cases}
	\begin{tikzpicture}
	\draw (-0.5,0) -- (0.5,0);
	\filldraw [gray] (-0.1,-0.1) -- (-0.1,0.1) -- (0.1,0.1) -- (0.1,-0.1);
	\end{tikzpicture}
	^\nu
	=
	\begin{tikzpicture}
	\draw [black](-0.5,0) -- (0.5,0);
	\end{tikzpicture}
	^\nu
	- \dfrac{1}{2}\;
	\begin{tikzpicture}
	\draw (-0.8,0) -- (-0.5,0);
	\draw (0.5,0) -- (0.8,0);
	\draw (-0.5,0) -- (-0.6,0.3)-- (-0.4,0.3);
	\draw (0.5,0) -- (0.6,0.3) -- (0.4,0.3);
	\fill [gray] (-0.5,0) -- (-0.6,0.3) -- (-0.4,0.3); 
	\fill [gray] (0.5,0) -- (0.6,0.3) -- (0.4,0.3); 
	\fill [white] (0.5,0) arc (0:180:0.5);
	\draw (0.5,0) arc (0:-180:0.5);
	\draw (0.5,0) arc (0:180:0.5);
	\filldraw [black] (0, 0.5) circle (2pt);
	\filldraw [white] (-0.1,-0.4) -- (0.1,-0.4) -- (0.1,-0.6) --(-0.1,-0.6)--cycle;
	\filldraw[black] (-0.1,-0.4) -- (0.1,-0.4) -- (0.1,-0.6) --(-0.1,-0.6)--cycle;
	\end{tikzpicture},
	\\[1ex]
	\begin{tikzpicture}
	\draw (-0.5,0) -- (0.5,0);
	\draw [dashed] (0,0) -- (0,0.5);
	\filldraw [gray] (0, 0) circle (3pt);
	\end{tikzpicture}
	=
	\begin{tikzpicture}
	\draw (-0.5,0) -- (0.5,0);
	\draw [dashed] (0,0) -- (0,0.5);
	\end{tikzpicture}
	+
 \begin{tikzpicture}
 \fill [gray] (0.6,0) -- (0.45,-0.3) -- (0.3,0); 
	\fill [gray] (-0.6,0) -- (-0.45,-0.3) -- (-0.3,0);
	\fill [gray] (0,0.6) -- (-0.2, 0.8) -- (-0.26, 0.55);
 \fill [white] (0.6,0) arc (0:180:0.6);
	\draw (-1,0) -- (1,0);
	\draw (0.6, 0) arc (0:180:0.6);
	\draw [dashed] (0,0.6) -- (0,1);
	\filldraw [black] (0, 0) circle (2pt); 
	\filldraw [black] (0.42, 0.42) circle (2pt);
	\filldraw [black] (-0.42, 0.42) circle (2pt);
	\end{tikzpicture}.
 \end{cases}
\end{equation}
The first equation allows us to determine the $2$-point function, while the second one is for the $3$-point
function. Dotted lines represent a vanishing incoming momentum, the decorations represent the functions we compute
\begin{gather*}
G(a,L):= \sum_{n\geq 0} \frac{1}{n!}\gamma_n(a) L^n
	\qquad \text{for} \quad
	\begin{tikzpicture}
		\draw (-.5,0) -- (.5,0);
		\filldraw [black] (0, 0) circle (2pt);
	\end{tikzpicture},
	\\
Y(a,L):=\sum_{n\geq 0} \frac{1}{n!} \upsilon_n(a) L^n
	\qquad \text{for} \quad
	\begin{tikzpicture}
		\draw (-.5,0) -- (.5,0);
		\fill [gray] (-.2,0) -- (0,.3) -- (.2,0); 
	\end{tikzpicture}
\end{gather*}
and the square appears once and only once in the diagrams for the derivative of the propagator
\begin{gather*}
K^\nu (a,L) := \partial^\nu \bigg(\frac{G(a,L)}{p^2}\bigg) = \frac{2p^\nu}{(p^2)^2} \sum_{n \geq 0} (\gamma_{n+1}-\gamma_n) \frac{L^n}{n!}
	\qquad \text{for} \quad
	\begin{tikzpicture}
		\draw (-.5,0) -- (.5,0);
		\fill [gray] (-.15,-.15) -- (-.15,.15) -- (.15,.15)--(.15,-.15); 
	\end{tikzpicture}.
\end{gather*}
In these equations, $a:=g^2/(4\pi)^3$ is an equivalent of the fine structure constant which hides irre\-le\-vant
$\pi$ factors, $g$ is the coupling constant, $\partial^\nu := {\partial}/{\partial p_\nu}$, and
 $L:=\log \big(p^2/\mu^2\big)$ is the logarithmic kinematic variable for a reference energy scale $\mu^2$.
These decorations and the functions they denote are always relative to the free propagator. They do not depend on
any regularisation parameter and satisfy the renormalisation group
equations
\begin{gather}
\partial_L G = (\gamma + \beta a\partial_a) G, \label{renormG}
\\[.5ex]
\partial_L Y = (\upsilon + \beta a\partial_a) Y, \label{renormY}
\end{gather}
where $\gamma (a)$ and $\upsilon (a)$ are the anomalous dimensions of the $2$-point and $3$-point function and~$\beta(a)$ is the beta-function of the model. In~\cite{BeRu20}, we established that $\beta = 2 \upsilon + 3
\gamma$, but we must point out that our convention for the function $\beta$ differs from the usual ones.
Since $\gamma_0$ and $\upsilon_0$ are left undetermined by the equations, we fix them to 1 as a normalisation
condition. Then equations \eqref{renormG} and~\eqref{renormY} impose the relations $\gamma_1=\gamma$ and
$\upsilon_1=\upsilon$ which are used to determine the renormalisation group functions $\gamma$ and
$\upsilon$.

Decorations can be further composed as Cauchy products of the $G$ and $Y$ series.
We can assign to each internal line an operator $\Ww$ given as a product of $G$ and~$Y$,
with its anomalous dimension $w$ given as
\begin{gather*}
 w= \#G\, \gamma + \#Y\, \upsilon,
\end{gather*}
and the renormalisation group equation
\begin{gather*}
\partial_L \Ww = (w + \beta a \partial_a) \Ww.
\end{gather*}
Here is a non-exhaustive list of operators $\Ww$ that we could consider
\begin{gather*}
 \Ww \in \big\{
\begin{tikzpicture}
		\draw (-.5,0) -- (.5,0);
		\filldraw [black] (0, 0) circle (2pt);
\end{tikzpicture}
,
\begin{tikzpicture}
		\draw (-.5,0) -- (.5,0);
		\fill [gray] (-.15,0) -- (0,.3) -- (.15,0); 
		\filldraw [black] (-.3, 0) circle (2pt);
		\filldraw [black] (.3, 0) circle (2pt);
\end{tikzpicture}
,
\begin{tikzpicture}
		\draw (-.5,0) -- (.5,0);
		\fill [gray] (-.5,0) -- (-.35,.3) -- (-.2,0); 
		\filldraw [black] (.1, 0) circle (2pt);
\end{tikzpicture}
,
\begin{tikzpicture}
		\draw (-.5,0) -- (.5,0);
		\fill [gray] (.5,0) -- (.35,.3) -- (.2,0); 
		\filldraw [black] (-.1, 0) circle (2pt);
\end{tikzpicture}
,
\begin{tikzpicture}
		\draw (-.5,0) -- (.5,0);
		\fill [gray] (.5,0) -- (.35,.3) -- (.2,0); 
		\fill [gray] (-.5,0) -- (-.35,.3) -- (-.2,0); 
		\filldraw [black] (0, 0) circle (2pt);
\end{tikzpicture}
,
\begin{tikzpicture}
		\draw (-.8,0) -- (.8,0);
		\fill [gray] (.8,0) -- (.65,.3) -- (.5,0); 
		\fill [gray] (-.8,0) -- (-.65,.3) -- (-.5,0); 
		\fill [gray] (-.15,0) -- (0,.3) -- (.15,0); 
		\filldraw [black] (.3, 0) circle (2pt);
		\filldraw [black] (-.3, 0) circle (2pt);
\end{tikzpicture}
 \big\}.
\end{gather*}

In equations~\eqref{WSDdiag}, two products appear, $\s$ (read ``Samekh'') and $\q$ (read ``Qof''), defined as
\begin{gather*}
\s (a,L):= YGY (a,L) = \sum_{n\geq 0} \frac 1{n!} s_n(a) {L^n}
	\qquad \text{for} \quad
	\begin{tikzpicture}
		\fill [gray] (-.7,0) -- (-.55,.3) -- (-.4,0); 
		\fill [gray] (.7,0) -- (.55,.3) -- (.4,0); 
		\draw (-.7,0) -- (.7,0);
		\filldraw [black] (0, 0) circle (2pt);
	\end{tikzpicture},
	\\
\q (a,L):=GYG (a,L) = \sum_{n\geq 0} \frac 1{n!} q_n(a) L^n
	\qquad \text{for} \quad
		\begin{tikzpicture}
		\fill [gray] (-.15,0) -- (0,.3) -- (.15,0); 
		\draw (-.7,0) -- (.7,0);
		\filldraw [black] (-.45, 0) circle (2pt);
		\filldraw [black] (.45, 0) circle (2pt);
	\end{tikzpicture}.
\end{gather*}
In particular, they satisfy
\begin{gather*}
\partial_L \s = (\gamma + 2\upsilon + \beta a\partial_a) \s,
 \qquad
\partial_L \q = (2\gamma + \upsilon + \beta a\partial_a) \q,
\end{gather*}
and we will write $s= 2\upsilon + \gamma$ and $q= 2 \gamma + \upsilon$.
This formalism was tested in~\cite{BeRu20} by showing that the renormalisation functions to order $a^2$ computed
with it matched with known results.

As is customary in resurgence literature, we work with a variable defined in the neighbourhood of infinity, that
is one which is proportional to $1/a$. We use the variable $r$ with a normalisation such that:
\begin{gather}\label{betar}
r\beta (r)= -1 + o\bigg(\frac1r\bigg).
\end{gather}
This means that we define $r = -{1}/({\beta_1 a})$ and the sign here is important because we want to keep this
trademark of asymptotic freedom. This normalisation ensures that the transseries terms in~the expansion of the
anomalous dimensions will contain $\exp(nr)$ factors with integer $n$, and the exponents we compute have
simpler expressions.

Our computations make use of the Mellin transform representation of graphs. It is obtained by replacing the $n_e$ propagators by full propagators with an additional $\exp( x L)$ factor. The evaluation of the
resulting modified graph gives a function of $n_e$ complex parameters, meromorphic with poles
on linear subspaces.

The usefulness of such a function stems from the relation
\begin{gather*}
L^n =  \partial_x^n {\rm e}^{xL} \big\vert_{x=0},
\end{gather*}
which permits to obtain any function of~$L$ through the action of an infinite order differential operator
on~$\exp(xL)$. The effect of the replacement of any propagator in a diagram can be obtained from the action of
this differential operator on (one of the parameters of) the Mellin transform, followed by putting the parameter to~0.
So, for example, the $G$ series is descri\-bed~by
\begin{gather*}
G_x := G(a,\partial_x) = \sum_{n\geq 0} \gamma_n(a) \frac{\partial_x^n}{n!}.
\end{gather*}
We assign also an operator $\Oo$ to the whole graph by multiplying the operators associated to all of the internal
lines $\Ww_i$, with a factor $1/r^l$ for a diagram with $l$ loops, so that the evaluation of the graph is just
by applying this operator on the Mellin transform of the graph. For example, for the graphs appearing in
equations~\eqref{WSDdiag}
\begin{gather*}
\Oo_{xy} =
	\begin{cases}
	\dfrac{1}{r} K^\nu_x \s_y & \text{for} \quad
		\begin{tikzpicture}
		\draw (-0.8,0) -- (-0.5,0);
		\draw (0.5,0) -- (0.8,0);
		\draw (-0.5,0) -- (-0.6,0.3)-- (-0.4,0.3);
		\draw (0.5,0) -- (0.6,0.3) -- (0.4,0.3);
		\fill [gray] (-0.5,0) -- (-0.6,0.3) -- (-0.4,0.3); 
		\fill [gray] (0.5,0) -- (0.6,0.3) -- (0.4,0.3); 
		\fill [white] (0.5,0) arc (0:180:0.5);
		\draw (0.5,0) arc (0:-180:0.5);
		\draw (0.5,0) arc (0:180:0.5);
		\filldraw [white] (0.1,-0.4) -- (-0.1,-0.4) -- (-0.1,-0.6) -- (0.1,-0.6)--cycle;
		\filldraw[black] (-0.1,-0.4) -- (0.1,-0.4) -- (0.1,-0.6) --(-0.1,-0.6)--cycle;
		\filldraw [black] (0,0.5) circle (2pt);
		\end{tikzpicture},
	\\[2ex]
	\dfrac{1}{r} \q_x \s_y & \text{for} \quad
		\begin{tikzpicture}
 		\fill [gray] (0.6,0) -- (0.45,-0.3) -- (0.3,0); 
		\fill [gray] (-0.6,0) -- (-0.45,-0.3) -- (-0.3,0);
		\fill [gray] (0,0.6) -- (-0.2, 0.8) -- (-0.26, 0.55);
 		\fill [white] (0.6,0) arc (0:180:0.6);
		\draw (-1,0) -- (1,0);
		\draw (0.6, 0) arc (0:180:0.6);
		\draw [dashed] (0,0.6) -- (0,1);
		\filldraw [black] (0, 0) circle (2pt); 
		\filldraw [black] (0.42, 0.42) circle (2pt);
		\filldraw [black] (-0.42, 0.42) circle (2pt);
		\end{tikzpicture}.
	\end{cases}
\end{gather*}
We do not keep $K^\nu$ in our operators, since it can be expressed through $G$ and its derivative with
respect to~$L$: $K^\nu_x$ can be traded for~$G_x$, with only a multiplication of the Mellin transform
by~$x-1$. In the one loop diagram we consider here, the propagator $K^\nu$ is the only one in a path between
the exterior legs of the diagram, so that it could be obtained through the derivation of~the whole diagram with
respect to the exterior momentum. In any case, we can just write $G$ instead of $K^\nu$.
We can assign to $\Oo$ an anomalous dimension $\gamma_{\Oo}$:
\begin{gather}\label{gammaO}
\gamma_\Oo \equiv \#G \, \gamma + \#Y\, \upsilon - l \beta,
\end{gather}
and write
\begin{gather*}
\partial_ L \Oo = ( \gamma_\Oo - \beta r \partial_r ) \Oo.
\end{gather*}
The $\beta$ function appears in equation~\eqref{gammaO} due to the $1/r^l$ factor in the definition of~$\Oo$.
It is a~combination of the anomalous dimensions $\gamma$ and $\upsilon$,
in our model
\begin{gather*}
\beta = 3\gamma + 2\upsilon,
\end{gather*}
such that for the graphs in equation~\eqref{WSDdiag}
\begin{gather*}
\gamma_\Oo =
\begin{cases}
-\gamma &\text{for} \quad \Oo_\gamma:=\dfrac1r G_x \s_y,
\\[1.5ex]
\phantom{-}\upsilon &\text{for} \quad \Oo_\upsilon:=\dfrac1r \q_x \s_y.
\end{cases}
\end{gather*}
This ensures that both sides of the WSD equations obey the same renormalisation group equations and is
instrumental in the proof we have given in~\cite{BeRu20} that the solutions of the WSD equations obey
renormalisation group equations.
The $\beta$ function is then the logarithmic derivative of the effective charge $r^{-1}Y^2 G^3$, which have
important combinatorial properties, see for example~\cite{BaKrUmYe09}.

From Ward--Schwinger--Dyson equations~\eqref{WSDdiag} we can extract equations for the anomalous dimensions:
\begin{gather*}
\gamma = (\gamma + \beta r\partial_r) \gamma - \frac{T_2}{2\beta_1}\Oo_\gamma H^\gamma,
\\
\upsilon = \frac{T_3}{\beta_1} \Oo_\upsilon H^\upsilon,
\end{gather*}
where $H^\gamma$ and $H^\upsilon$ are functions associated to the two graphs through Mellin transforms.

\section{Singularity structure of the Mellin transform}
\subsection{General properties}
The poles of the Mellin transforms $H^\gamma$ and $H^\upsilon$ give the dominant contributions in the evaluation
of the anomalous dimensions. These two functions are given by
\begin{gather*}
H^\gamma(x,y) = \frac{\Gamma(1-x-y)\Gamma(2+x)\Gamma(2+y)}{\Gamma(4+x+y)\Gamma(1-x)\Gamma(1-y)},
\\[.5ex]
H^\upsilon(x,y)= \frac{\Gamma(1-x-y)\Gamma(1+x)\Gamma(2+y)}{\Gamma(3+x+y)\Gamma(2-x)\Gamma(1-y)},
\end{gather*}
and they have poles when the argument of one of the $\Gamma$ function in their numerators is a
negative integer or zero. In terms of natural integers $n$, $n'$, $n''$, we have therefore poles
on the line with equations
\begin{gather*}
\begin{cases}
2+x = -n ,\\
2+y = -n' ,\\
1-x-y = -n'' ,
\end{cases}
\qquad
\text{and}
\qquad
\begin{cases}
1+x = -n ,\\
2+y = -n' ,\\
1-x-y = -n'' ,
\end{cases}
\end{gather*}
respectively for $H^\gamma$ and~$H^\upsilon$.
\begin{figure}[h]
\centering
\includegraphics[scale =.52]{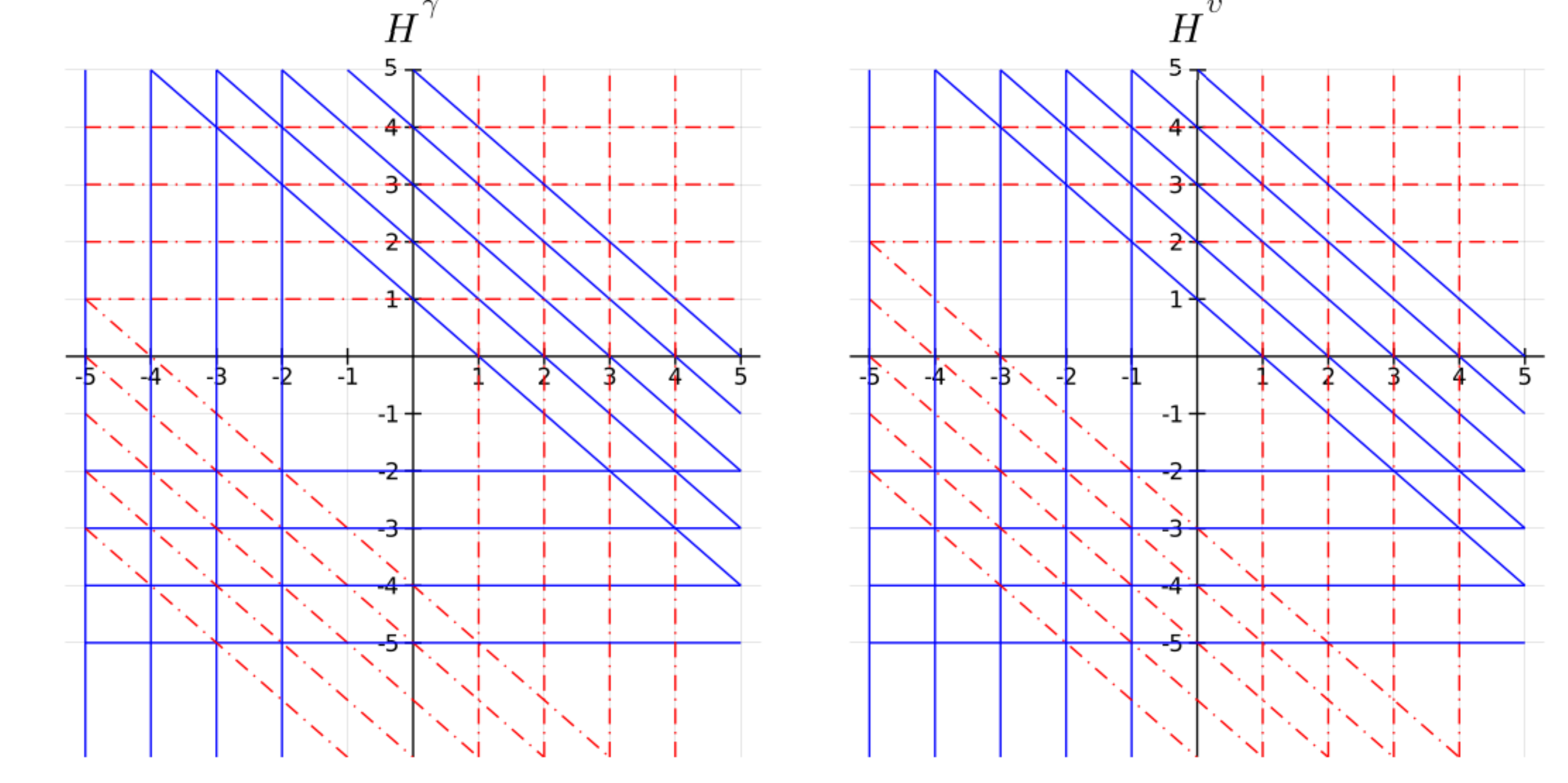}
\caption{Singularity structure of the characteristic functions $H$:
the continuous blue lines represent the poles, the dot-dashed red ones the zeroes.}
\label{figura}
\end{figure}

The singularity structures of these functions are represented in Figure~\ref{figura}. We can also infer from these
plots the properties of the residue along any pole line. Consider one of them, whenever a~zero line crosses
it, the residue get a zero, unless another pole line goes through the intersection point. One can see that a
zero line is present at each intersections of two pole lines, so that the residues never get poles of their own.
In both cases, the singularity represented by the blue line $1-x-y=0$ is the closest one to the origin.
For any $H$ function we write the decomposition
\begin{gather} \label{Hdecomp}
	H(x,y) = \sum_k \frac{h_k(x,y)}{k-x-y} + \frac{h'_k(y)}{k+x} + \frac{h''_k(x)}{k+y} +
		\sum_{n,m} \tilde{h}_{n,m} x^n y^m,
\end{gather}
where the index sum is intentionally left unspecified because it depends on the particular choice of $H$ and a
priori there should be four different summations but we did not want to weight the notation too much. This
description separates the poles of different kind whose residues are $h_k$, $h'_k$, $h''_k$ and an analytic part
described by~$\tilde{h}_{n,m}$. Obviously if $H$ is symmetric for $x\leftrightarrow y$ then $h'_k=h''_k$, but it
is not the case for~$H^\upsilon$.
We give here their values for small~$k$:
\begin{center}\def\arraystretch{1.3}
\begin{tabular}{c|c|c|c|c}
\hline
& $k$	&	$h_k$ & $h'_k$ &	$h''_k$
\\
\hline\rule{0ex}{4ex}%
$H^\gamma$& $1$ &	 $ \dfrac{xy(2+xy)}{4!}$ & --- & ---
\\[1.5ex]
& $2$ &$ \dfrac{xy(3+xy)(1-xy)}{5!}$ & $\dfrac{(y-2)(y-1)}{2} $ & $\dfrac{(x-2)(x-1)}{2} $
\\[1.5ex]
& $3$ &$ \dfrac{xy (xy-2)^2 (xy+4)}{6! 2} $ & $\dfrac{\big(y^2-1\big)(2-y)(3-y)}{3!} $ & $\dfrac{\big(x^2-1\big)(2-x)(3-x)}{3!}$
\\[1.5ex]
\hline\rule{0ex}{4ex}%
$H^\upsilon$	& $1$ &$ \dfrac{x(1+y)}{3!}$ & $\dfrac{1-y}{2}$ & ---
\\[1.5ex]
& $2$ &$ \dfrac{xy(1+y)(1-x)}{4!}$ & $\dfrac{\big(1-y^2\big)(y-2)}{3!} $ & $\dfrac{2-x}{2} $
\\[1.5ex]
\hline
\end{tabular}
\end{center}
The decompositions~\eqref{Hdecomp} of the functions~$H(x,y)$ give corresponding decompositions of the $\Oo H$
constructs. The poles for $ x + y =k$ give rise to functions~$E_k$ in this decompositions. For the poles
depending on a single variable, the corresponding term can be further decomposed as a~product of factors
respectively associated to the variables $x$ and~$y$: this part of~$H(x,y)$ is the product of two
functions depending on a single variable and the action of $\Oo$ on it will give the product of the action
of~$\Oo_x$ on the function of~$x$ and the action of~$\Oo_y$ on the function of~$y$. The~factor corresponding
to a single pole in a single variable will be called~$F_k$.
Functions $E_k$ and~$F_k$, were first considered in~\cite{Be10a} and were fundamental
tools in~\cite{BeSc12} and in~\cite{BeCl13}.

\subsection[Fk functions]{$\boldsymbol{F_k}$ functions}
These functions capture the contribution to anomalous dimensions due to $k+x=0$ or $k+y=0$ poles of the characteristic functions $H$.
For $\Ww$ in $\{G,\s,\q\}$ and the corresponding $w$ in $\{\gamma, s, q \}$, we define
\begin{gather*}
F^w_k:= \Ww_x \bigg(\frac{x}{k+x}\bigg)=\Ww_x \bigg(\sum_{n\geq 0} \frac{(-1)^n x^{n+1}}{k^{n+1}}\bigg)
		=- \sum_{m\geq 1} \frac{(-1)^m w_m }{k^m}.
\end{gather*}
We could have defined $F^w_k$ as the action of $\Ww$ on the rational function ${1}/({k+x})$, but this would
produce terms which have a constant part and this would bring some difficulties with the Borel transforms.

It follows immediately from the definition that
\begin{gather} \label{FkRG}
\partial_L \Ww \bigg(\frac{x}{k+x}\bigg) = -k F^w_k + w.
\end{gather}
Indeed, we can write
\begin{align*}
\partial_L \Ww \sum_{n\geq 0} (-1)^n \frac{x^{n+1}}{k^{n+1}}
&= ( w -\beta r \partial_r ) \sum_{n\geq 0} (-1)^n \frac{w_{n+1}}{k^{n+1}}
=\sum_{n\geq 0} (-1)^n \frac{w_{n+2}}{k^{n+1}}
\\
&= k \sum_{n\geq 2} (-1)^n \frac{w_n}{k^n}=-k\bigg( F^w_k - \frac{w}{k}\bigg).
\end{align*}
The generic equation~\eqref{FkRG} can be specialised to the three different kinds of $F_k$, associated to the
three kinds of propagator like operators appearing in the Ward--Schwinger--Dyson equations. The resulting
equations allow to compute the corresponding $F_k$:
\begin{gather*}
(\gamma-\beta r \partial_r)F_k^\gamma = \gamma - kF_k^\gamma,
\\[.5ex]
(s-\beta r \partial_r)F_k^s = s - kF_k^s,
\\[.5ex]
(q-\beta r \partial_r)F_k^q = q - kF_k^q.
\end{gather*}
There is potentially an infinity of $F_k$ terms which can contribute to the WSD equations, but in our
computations, we will need only a finite number of them. The first one in importance is~$F_1^q$ but we also
have all the $F_k$ with a constant term in the residue of the corresponding pole.
These cases can be read in Figure~\ref{figura} since they correspond to the poles parallel to an axis which are
not crossed by zeroes at their intersection with the other axis. In the case of $H^\gamma$, since it is symmetric
by $x\leftrightarrow y$, it suffices to describe the poles along one direction. These poles come from
$\Gamma(2+x)$ and thus start for $x=-2$ and end when a zero is crossed for $y=0$, produced by the term
$1/\Gamma(4+x+y)$, therefore for $x=-4$ and lower. In the end, we have contributions from two values of
$k$, 2 and 3.
In the case of $H^\upsilon$, we do not have this symmetry and we must distinguish the two directions: along $x$,
the poles start at $x=-1$, from $\Gamma(1+x)$, and hit a zero at $x=-3$, from $1/\Gamma(3+x+y)$ taken at
$y=0$, leaving the two poles at $-1$ and $-2$.
Along the $y$ direction, the important poles are determined by $\Gamma(2+y)$ for the start of the poles and get a~zero in the residue from the same factor than for the $x$ poles, leaving only one pole for $y=-2$.
In the equation for the anomalous dimension $\gamma$ there will be contributions from $F_k^s$ and~$F_k^\gamma$,
while for the one for the anomalous dimension $\upsilon$ there will be contributions from $F_k^s$ and~$F_k^q$.

\subsection[Ek functions]{$\boldsymbol{E_k}$ functions}
The $E_k$ functions capture the contribution to anomalous dimensions due to $k-x-y=0$ poles of the characteristic functions $H$.
Differently from $\boldsymbol{F}_k$ functions, they are not associated to a single propagator but to the whole diagram. We define them through the action of the operators~$\Oo_{xy}$:
\begin{gather}\label{defE_k}
\boldsymbol{E}_k = \Oo_{xy} \frac{\boldsymbol{h}_k(x,y)}{k-x-y}.
\end{gather}
In this article we will work with the functions $E^\gamma_k$ and $E^\upsilon_k$ given by
\begin{gather}
E^\gamma_k = \frac{G_x \s_y }{r} \frac{h^\gamma_k( x,y ) }{k-x-y}, \label{defE_k1}
\\
E^\upsilon_k = \frac{\q_x \s_y}{r} \frac{h^\upsilon_k( x,y ) }{k-x-y}. \label{defE_k2}
\end{gather}
As remarked already in~\cite{Be10a}, the relation~\eqref{defE_k} can also be written with exchanged places of
derivatives and variables
\begin{gather*}
\boldsymbol{E}_k = \frac{\boldsymbol{h}_k (\partial_1, \partial_2)}{k-\partial_1-\partial_2} \Oo (L_1,L_2),
\end{gather*}
where $\partial_i$ stands for $\partial_{L_i}$ and as usual, the variables are set to zero after all
differentiations are evaluated.
Inverting $k-\partial_1 - \partial_2$ is challenging, but when acting on $\Oo(L_1,L_2)$, it can be brought to
a form that only refers to the $r$ variable and therefore act similarly on the function~$E_k$. We~can also
see that an arbitrary power of $\partial_1 + \partial_2$ followed by the evaluation at $L_1=L_2=0$ can be
replaced by first evaluating $L_1=L_2=L$, applying the same power of $\partial_L$ and finally put $L=0$.
We can then write, with $\Oo' = \boldsymbol{h}_k(\partial_1,\partial_2) \Oo$,
\begin{gather*}
	\frac{1}{k-\partial_1-\partial_2} \Oo' = \frac 1 k \sum_n \frac{1}{k^n} \partial_L^n \Oo'
	= \frac 1 k \sum_n \frac{1}{k^n} (\gamma_\Oo - \beta r \partial_r)^n \Oo'
	= \frac{1}{k-\gamma_\Oo + \beta r \partial_r} \Oo',
\end{gather*}
which ultimately brings the equation
\begin{gather*}
(k - \gamma_\Oo + \beta r \partial_r) \boldsymbol{E}_k = \boldsymbol{h}_k (\partial_1, \partial_2) \Oo(L_1, L_2).
\end{gather*}
The rather formal definitions~\eqref{defE_k1} and \eqref{defE_k2} can be converted to the following
equations
\begin{gather*}
(k+\gamma + \beta r \partial_r) E^\gamma_k = h^\gamma_k(\partial_1,\partial_2) \frac{G(L_1) \s(L_2)}{r},
\\[.5ex]
(k-\upsilon+\beta r\partial_r)E^\upsilon_k=h^\upsilon_k(\partial_1,\partial_2)\frac{\q(L_1) \s(L_2)}{r}.
\end{gather*}
One must remark that it is the inclusion of the $1/r$ factor in the definition of these $E_k$ that brings the
simplification of the anomalous dimensions $\gamma_\Oo$ to $-\gamma$ and $\upsilon$.

Using the functions $E_k$ and~$F_k$, the contribution of the diagrams can be written as
\begin{gather}\label{OH}
\Oo H = \frac{c}{r} + \sum_k E_k + \frac{1}{r} \sum_w \sum_{k }
\mathfrak{z}^k_w F^w_k + R,
\end{gather}
with $c$ the constant giving the leading term and $R$ collecting all other possible terms.
Notice that ${E}_k$ functions include a factor $1/r$ while the ${F}_k$ do not. This explains that the two
sums do not come with the same $1/r$ factor.
The $F_k^w$ come with a factor $\mathfrak{z}^k_w$, which is just a number which is given by
\begin{gather*}
\mathfrak{z}^k_w = - \frac1k h^w_k(0),
\end{gather*}
where $h^w_k(0)$ denotes the constant term of either $h'_k(y)$ or $h''_k(x)$ according to which Mellin variable is associated to the line and the prefactor $-{1}/{k}$ comes from the identity
\begin{gather*}
\frac{1}{k+x} = \frac{1}{k} \bigg(1 -\frac{x}{k+x} \bigg).
\end{gather*}
We could have kept other terms of the residue, but they would not contribute to the exponents.

\subsection[R function]{$\boldsymbol R$ function}
Our computations presume that the dominant contributions come from the poles of the Mellin transform and more
specifically from the poles near the origin. We therefore need some way of bounding the contributions coming
from the remainder of the Mellin transform, once a finite number of poles has been subtracted.

The solution is not easy, since for the terms with exponential factors, all derivatives with respect to $L$
of the propagator corrections are now of the same order. In a first attempt to find the corrections to the
asymptotic behaviour of the series for the Wess--Zumino model~\cite{BeCl13}, a solution could be devised by
using the conjecturally exact expansion of the Mellin transform as sum of the pole contributions, when using a
particular extension of the residues to the whole~$(x,y)$ plane. This approach is however limited by the
appearance of multizeta values with high depth, which rapidly go beyond the cases with known reductions.

In a following work~\cite{BeCl14}, we could find a much easier solution. It used a transformation of the
propagator which can be written as
\begin{gather*}
G(L) = \sum_n \frac 1{n!} \gamma_n L^n \longrightarrow \check G(\lambda) = \sum_n \gamma_n
\lambda^{-n-1}.
\end{gather*}
The relation between $G$ and its transform looks like $G$ is the Borel transform of this new function~$\check G$, but it is not a proper interpretation, since the natural product for $G$ is not a~convolution
product. Nevertheless, the derivation with respect to $L$ becomes the multiplication by $\lambda$ after this
transform, so that it is easy to convert the renormalisation group equation for $G$ in an~equation for $\check
G$. In this equation, a term with an $\exp( k r)$ factor in $\check G$ gets multiplied by
$\lambda - k$, so that $\check G$ gets a singularity for $\lambda=k$.

On the other side, the pairing of $G$ with the Mellin transform is straightforward in this form. We just have
to make the sum of the products of the $x^n$ terms in $H$ and the $\lambda^{-n-1}$ terms in~$\check G$
which is just the residue of $H(x,y) \check G(x) $ at $x=0$. This can be conveniently expressed as a~contour
integral around the origin. If $\check G$ depends on $\exp(k r)$ and has therefore singularities for $x
=k$, the contour integral will involve also evaluation of $H$ or its derivatives at
$x=k$: this certainly does not work out if $H$ has a pole at this point, so that this computation can only be
done after subtracting a number of poles of $H$, but the upshot is that one can obtain in this way the
contribution from the remainder of $H$ in a form which contains sufficiently many $1/r$ factor to not have any
influence on the exponents we compute.

This construction must also be applied for the $\s$ and $\q$ products, which must be
directly obtained from their respective renormalisation group equations, since the pointwise product in~the
variable $L$ has no easy equivalent in the variable~$\lambda$. The full development of this formalism is
certainly complex, since the Mellin transform must be evaluated not only at the origin, but also, after
suitable subtractions, at integer points. Nevertheless, for the sake of our limited ambition in this work, it is
possible to consider that the rest function which accounts for all but linear terms in $h'_k$ and $h''_k$ and the
regular part of the $H$ function is controlled.
We can characterise it as being
\begin{gather*}
R = \oint\!\! \oint H_{\rm reg} \check{\Oo},
\end{gather*}
where $H_{\rm reg}$ is the function $H$ with subtracted polar parts. This subtraction unfortunately cannot be put in
the form of a canonical projection as would be possible for $H$ a function of a single variable.

\section{Trans-series corrections}
\subsection{Results}
From the resurgent point of view the situation is quite intricate, but there are things that we can easily establish.
First of all, from the dominant term in~$r\beta(r)$ in equation~\eqref{betar}, one sees that there is a
dominant $(k+\partial_r)F_k$ term in the equations for the $F_k$ and $(k-\partial_r) E_k$ in the ones for
the~$E_k$. We therefore see that $F_k$ can be modified by a term proportional to ${\rm e}^{-kr}$
and~$E_k$ by a term proportional to ${\rm e}^{kr}$. This is however not sufficient to characterise these terms.
The~next term in an expansion in~$1/r$ cannot be compensated by the derivation of a series in powers of~$r^{-1}$,
which have a derivative starting with $r^{-2}$, so that one has to multiply such terms by a power of~$r$,
generically with a non-integer exponent which will be the dominant one in the exponential terms in order to
cancel the terms of order~$1/r$.

The purpose of this section is to show how to compute the values of these dominant exponents for the corrections
proportional to ${\rm e}^r$ and ${\rm e}^{-r}$ for the anomalous dimension $\gamma$ and $\upsilon$. The situation appears
more complex than in the Wess--Zumino model, where there is only one $E_k$ and one $F_k$ for each positive
integer~$k$ and all exponents have been computed in~\cite{BeCl14}.
Talking about the first trans-series corrections means to talk about the closest singularities in the Borel
plane through their relations to alien derivatives. In turn, these singularities control the asymptotic
behaviour of the perturbative series.
Terms that are proportional to ${\rm e}^r$ are linked to the singularities at~$-1$ in the Borel plane while the ${\rm e}^{-r}$
terms are linked to the singularities at~1. We will use the notation that $[k]$ indicates the part of a
function which has a factor ${\rm e}^{kr}$, so that $[0]$ indicates the classical part and we will compute the
exponents for the $[1]$ and $[-1]$ parts, corresponding respectively to the singularities in $-1$ and $1$
of the Borel transform.

The results of this section are expressed in terms of three quantities $g$, $u$ and~$b$ appearing in
the first orders of the renormalisation group functions as
\begin{gather*}
\gamma [0] = \frac{g}{r} + O\bigg(\frac1{r^2}\bigg) = -\frac{T_2}{12\beta_1} \frac{1}{r} +O\bigg(\frac1{r^2}\bigg),
\\[.5ex]
\upsilon [0] = \frac{u}{r} + O\bigg(\frac1{r^2}\bigg) = \frac{T_3}{2\beta_1}\frac1r+O\bigg(\frac1{r^2}\bigg),
\\
\beta [0] r = (3\gamma [0] + 2 \upsilon [0]) r = -1 +\frac{b}{r} +O\bigg(\frac1{r^2}\bigg),
\end{gather*}
with
\begin{gather*}
\beta_1 = \frac{T_2}4 - T_3,
\\[.5ex]
b = \frac{\beta_2}{\beta_1^2} = \frac{1}{\beta_1^2} \bigg(\frac{11}{24}T_2T_3 - \frac{11}{144} T_2^2 -\frac{3}{4}T_3^2-\frac{1}{2}T_5\bigg).
\end{gather*}
These expressions appear different from previously published results, notably \cite{Gra2015}, due to our
conventions:
we write the renormalisation group equations with $\beta(r)$ as the variation of $\log r$. This shifts the
powers of $g$ and introduces a factor of $2$ in $\beta$.

The trans-series corrections will come with dominant powers of $r$. We will name their expo\-nents $\eta$, $\theta$ and the pair of conjugated numbers $\lambda^\pm$ as follows: $\gamma[1]$ is proportional to $r^\eta {\rm e}^r $; $\upsilon[1]$~to~$r^\theta {\rm e}^r
$; both $\gamma[-1]$ and $\upsilon[-1]$ are dominated by the two terms $r^{\lambda^\pm} {\rm e}^{-r}
$, with a definite relation between the dominant terms in $\gamma[-1]$ and~$\upsilon[-1]$.
Our results are summarized by
\begin{gather*}
\eta = g +b, \qquad
\theta= b-\frac23 u, \qquad
\lambda^{\pm} = -2g -b \pm |3g| \sqrt{1+\frac{4u}{3g}}.
\end{gather*}
Remarkably the difference
\begin{gather*}
\eta -\theta = \frac13
\end{gather*}
does not depend on the Casimir operators~$T_i$, apart in the case, where $\beta_1$ vanishes and our change of
variable is ill-defined.

We were surprised to find that
$\lambda^\pm$ is algebraic and even complex in general. Real asymptotic behaviors can only be obtained by
combining two conjugate terms involving $\lambda^+$ and~$\lambda^-$.

Let us show how they are calculated.

\subsection{Preparatory steps}

We start again from the WSD equations
\begin{gather*}
\gamma = (\gamma +\beta r \partial_r ) \gamma - \frac{T_2}{2\beta_1} \Oo_\gamma H^\gamma,
\\
\upsilon = \frac{T_3}{\beta_1} \Oo_\upsilon H^\upsilon,							
\end{gather*}
and rewrite them using equation~\eqref{OH} as
\begin{gather}
\gamma = (\gamma +\beta r \partial_r ) \gamma - \frac{T_2}{2\beta_1} \bigg( \frac{1}{6r} + \sum_k E^\gamma_k +\frac{1}{r} \bigg ({-}\frac{1}{2} \big(F^\gamma_2 + F^s_2\big) +\frac{1}{3} \big(F^\gamma_3 + F^s_3\big) \bigg)\bigg),
\label{WSDgammaultima}
\\
\upsilon = \frac{T_3}{\beta_1} \bigg(\frac{1}{2r} + \sum_k E^\upsilon_k +\frac{1}{r}
\bigg({-}\frac{1}{2} F_1^q +\frac{1}{6} F^q_2 -\frac{1}{2} F^s_2 \bigg)\bigg),
\label{WSDupsilonultima}
\end{gather}
while neglecting the $R$ terms.

For both ${\rm e}^r$ and ${\rm e}^{-r}$ trans-series order we can neglect $E_k$ for $k\geq 2$. This occurs because these functions satisfy the equations
\begin{gather*}
(k+\gamma + \beta r \partial_r) E^\gamma_k = h^\gamma_k(\partial_1,\partial_2) \frac{G(L_1) \s(L_2)}{r},
\\[.5ex]
(k- \upsilon + \beta r \partial_r) E^\upsilon_k = h^\upsilon_k(\partial_1,\partial_2) \frac{\q(L_1) \s(L_2)}{r},
\end{gather*}
and if we denote the descending powers by
\begin{gather*}
N^{\underline{n}} = N (N-1) \cdots (N-n+1)
\end{gather*}
we have
\begin{gather*}
h_k^\gamma (x,y) = \frac{(-1)^{k-1}}{(k-1)!\Gamma(4+k)} (2+x)^{\underline{k+2}} (2+y)^{\underline{k+2}},
\\
h_k^\upsilon (x,y) = \frac{(-1)^{k-1}}{(k-1)!\Gamma(3+k)} (1+y)^{\underline{k}} x^{\underline{k}}.
\end{gather*}
The lowest degree monomial will be $xy$ for these $h_k$ except for $h^\upsilon_1$, where it is just $x$.
Higher orders in $x$ and~$y$ for $h_k$ extract terms with higher order in~$L$ in the propagators and these
terms also begin at higher order in~$1/r$. Indeed, since we have that $w_{n+1}= (w - \beta r \partial_r ) w_n$,
$w_{n+1}$ is of one order higher than~$w_n$. Since the equations for the $E_k$ with $k$ larger than two
are not resonant, they are of the same order than the term produced by $h_k$, which is at least two order smaller
than $\gamma$ or~$\upsilon$, unable to modify the exponents. At this stage, we only have to consider~$E_1^\gamma$ and $E^\upsilon_1$. This simplification might not be true at higher trans-series order. From now on
let us forget the subscript $1$.

Furthermore, the $F_k$ functions do not contribute to the dominant exponents for trans-series term proportional
to ${\rm e}^r$, as can be seen from the computation of $F_k[1]$:
\begin{gather}\label{F[1]}
\bigl ((k+w -\beta r \partial_r) F^w_k \bigr) [1] = w[1].
\end{gather}
The left hand side can be expanded as
\begin{gather*}
\bigl((k+w -\beta r \partial_r) F^w_k\bigr) [1]
= (k+w -\beta r \partial_r)[0] F^w_k [1] + (k+w -\beta r \partial_r) [1] F^w_k [0].
\end{gather*}
If we parametrise $F^w_k [1]$ as follows
\begin{gather*}
F^w_k [1] = {\rm e}^r r^{f^w_k}c^w_k (1 +\cdots ),
\end{gather*}
we have
\begin{gather*}
\partial_r F^w_k [1] = \bigg(1 + \frac{f^w_k}{r} + \mathcal{O}\big(r^{-2}\big)\bigg) F^w_k [1].
\end{gather*}
Using that $F^w_k[0]$ is at least of order 1 in $1/r$, the dominant term of equation~\eqref{F[1]} gives that
\begin{gather*}
F^w_k [1] \sim \frac{1}{k+1} w[1],
\end{gather*}
which can be specialised to the following dominant behaviours
\begin{gather*}
F_k^\gamma [1] \sim \frac{1}{k+1} \gamma [1],
\\
F_k^s [1] \sim \frac{1}{k+1} s[1]= \frac{1}{k+1} (\gamma[1] + 2\upsilon[1]),
\\
F_k^q [1] \sim \frac{1}{k+1} q[1]= \frac{1}{k+1} (2\gamma[1] + \upsilon[1]).
\end{gather*}
Since in equations~\eqref{WSDgammaultima} and~\eqref{WSDupsilonultima} they appear with a prefactor of $r^{-1}$ they will always be subdominant and for our purpose negligible.
The dominant contributions proportional to ${\rm e}^r$ come from $E[1]$.

The opposite situation occurs for the ${\rm e}^{-r}$ terms. The $E[-1]$ do not contribute.
It suffices to consider the equations
\begin{gather*}
\big((1 +\gamma + \beta r \partial_r)E^\gamma\big) [-1]
= \frac{1}{12r} (\gamma s) [-1],
\\
\big((1 - \upsilon + \beta r \partial_r) E^\upsilon \big) [-1] = \frac{1}{6r} q[-1] , 
\end{gather*}
and observe that the left hand side is dominated by $(k+1)E_k [-1]$.
This happens because
$E_k [0] \in r^{-2}\mathbb{C}[[1/r]]$, since $\Oo$ contains a factor of $r^{-1}$ and $h_k$ do not contain constant terms; also
\begin{gather*}
\beta r \partial_r E[-1] \sim E [-1]
\end{gather*}
because the factor $-1$ coming from ${\rm e}^{-r}$ compensates with $\beta[0] r = -1 +\cdots$.

This means that $E^\gamma[-1]$ will always be subdominant. Any contribution are suppressed by two factors of $r$: one from the original equation and one from order $[0]$ series. So even if there was resonance, which in fact occur, the shift is sufficient to neglect these terms. The situation is, a priori, different for $E^\upsilon[-1]$. In fact due to the exceptional right hand side for $E^\upsilon$, where no order $[0]$ series appear, $E^\upsilon$ might contribute. Nevertheless we will see in Section~\ref{F-1} that $F_1^q[-1]$ is resonant and will dominate over $E^\upsilon$.

\subsection[Exponents of the leading singularities at -1]{Exponents of the leading singularities at $\boldsymbol{-1}$}
We take the following form for the functions $E^\gamma$ at the ${\rm e}^r$ level
\begin{gather*}
E^\gamma [1] = {\rm e}^r r^\epsilon c_\epsilon (1+\cdots) ,
\\
E^\upsilon [1] = {\rm e}^r r^{\bar{\epsilon}} c_{\bar{\epsilon}} (1+\cdots).
\end{gather*}
In equation~\eqref{WSDgammaultima}, the first term on the right hand side has the following leading contribution
\begin{gather*}
\bigl ((\gamma + \beta r \partial_r)\gamma \bigr) [1] \sim - \gamma[1]
\end{gather*}
which generates a factor of $2$ with the $\gamma[1]$ of the left hand side. Putting this in
equation~\eqref{WSDupsilonultima} we obtain the dominant terms
\begin{gather}
\gamma[1] \sim 3g E^\gamma_1 [1], \label{gfromE}
\\[.5ex]
\upsilon [1] \sim 2u E^\upsilon_1 [1]. \label{ufromE}
\end{gather}
Then $\epsilon$ and $\bar{\epsilon}$ can be obtained from the equations for $E^\gamma$ and~$E^\upsilon$
\begin{gather*}
\bigl((1+\gamma +\beta r \partial_r) E^\gamma \bigr) [1] \sim - \frac{1}{12r}(\gamma s)[1] ,
\\[.5ex]
\bigl((1-\upsilon +\beta r \partial_r) E^\upsilon \bigr)[1] \sim - \frac{1}{6r} q[1],
\end{gather*}
where we have neglected higher order terms on the right hand side. At this transseries order we have resonance:
the highest order terms in the left hand side cancel exactly due to $\beta [0] r = -1+ O(1/r)$. No contribution from
$E^\gamma[0]$ or $E^\upsilon[0]$ occurs because $E^\gamma[0]\in r^{-3}\mathbb{C}[[1/r]]$ and $E^\upsilon[0] \in
r^{-2}\mathbb{C}[[1/r]]$. We are left with
\begin{gather*}
\frac{g+b-\epsilon}{r} E^\gamma [1] \sim 0,
\\
\frac{-u +b -\bar{\epsilon}}{r} E^\upsilon [1] \sim -\frac{1}{6r} q[1].
\end{gather*}
In the first one of these equations, the right hand side is negligible, because it contains either~$\gamma[0]$ or
$s[0]$ which give an additional $1/r$ factor. In the last one, we use $q[1]= 2\gamma[1] + \upsilon[1]$ and
the relation~\eqref{ufromE} to end up with
\begin{gather*}
\epsilon = g + b,
\\
\bar{\epsilon} = - \frac23 u + b
\end{gather*}
in order to have non trivial solutions for $E^\gamma$ and $E^\upsilon$. Then, using the relations
\eqref{gfromE} and~\eqref{ufromE} shows that the dominant exponents in $\gamma[1]$ and~$\upsilon[1]$ are the
ones announced before.

\subsection[Exponents of the leading singularities at 1]
{Exponents of the leading singularities at $\boldsymbol 1$}\label{F-1}

Here the situation is less straightforward, with different important terms in the equations for $\gamma$ and
$\upsilon$ and interactions between them.

For $\gamma [-1]$, the leading resonance comes from the term $(\gamma + \beta r \partial_r) \gamma$. The $F_2$
and $F_3$ terms, even if they are not resonant, give contributions of the order $\gamma[-1]/r$ which cannot
be neglected.
The first subdominant terms come from $\beta[-1]$ and $\gamma[0]$. In particular equation~\eqref{WSDgammaultima}
becomes, when regrouping all $g\gamma[1]$ terms
\begin{gather*}
(g+ \lambda +b) \gamma[-1] + 2g \upsilon [-1] \sim 6g \bigg ({-}\frac12 \big( F_2^\gamma + F_2^s\big)[-1] +\frac13 \big(F_3^\gamma + F_3^s\big)[-1] \bigg).
\end{gather*}
The right hand side can be further simplified using
\begin{gather*}
F^w_k [-1] \sim \frac{1}{k-1} w[-1].
\end{gather*}
We therefore obtain that
\begin{gather*}
(g+ \lambda +b) \gamma[-1] + 2g \upsilon [-1] \sim -4g ( \gamma [-1] + \upsilon[-1] ),
\end{gather*}
giving
\begin{gather}
( 5g + \lambda +b ) \gamma[-1] \sim -6g \upsilon[-1] . \label{I}
\end{gather}
If $\upsilon[1]$ had a smaller exponent than $\gamma[1]$, this would give an equation for $\lambda$, but
this is not the case.

In the expression for $\upsilon [-1]$, $F_1$ stands out because it is resonant, so it adumbrates the other~$F_k$.
Indeed, in the equation for $F_1^q$
\begin{gather*}
\big((1 +q -\beta r \partial_r)\big) F_1^q [-1] = q[-1]
\end{gather*}
the highest terms cancel, so if we call $\varphi$ the dominant exponent for $F^q_1$, we have
\begin{gather}
\frac1r (q_1+b+\varphi) F_1^q [-1] \sim 2\gamma[-1] + \upsilon[-1], \label{II}
\end{gather}
with $q_1 = u + 2 g$ the first coefficient in~$q$.
The other $F_k$ do not have these cancellations, so that they give contributions smaller by a factor $1/r$ to
$\upsilon[-1]$ in equation~\eqref{WSDupsilonultima}, which at the approximation level we use, gives a further
relation
\begin{gather}
\upsilon[-1] \sim - \frac{u}{r} F_1^q [-1]. \label{III}
\end{gather}

Putting together equations~\eqref{I},~\eqref{II} and~\eqref{III}, we get an equation for the dominant power~$\lambda$:
\begin{gather*}
q_1+b+\varphi = -u (2\chi +1),
\end{gather*}
where
\begin{gather*}
\chi = \frac{-6g}{5g+\lambda+b}
\end{gather*}
is the proportionality factor between $\gamma[-1]$ and~$\upsilon[-1]$ from equation~\eqref{I}.
Equation~\eqref{II} further shows that $\varphi = \lambda + 1$ and we will use that, with our rescaling, the
first coefficient of~$\beta$ is $-1$ so that $1 = -3 g - 2 u$.
Finally $\lambda$ satisfies
\begin{gather*}
\lambda^2 + 2 ( 2g +b) \lambda + b^2 +4gb - 5g^2 -12 ug = 0 ,
\end{gather*}
with solutions
\begin{gather*}
\lambda^\pm = -2g -b \pm |3g| \sqrt{1 +\frac{4u}{3g}}.
\end{gather*}
The fact that it is algebraic is already strange, but for generic $\phi^3$ models the argument of the square root
is even negative, thus giving imaginary numbers. For example for the one-component case where $T_2=T_3=T_5=1$, its
value is $\lambda^\pm = \frac{107}{81} \pm \frac{{\rm i}\sqrt{7}}{3}$.

This result is quite surprising, since we expected that the exponents would remain rational, since they
only depend on the rational numbers $b$, $g$ and~$u$. In Section~\ref{alien}, we recalled how, through
the use of alien derivatives, a link can be established between deformations of the solution involving
exponential factors and the asymptotic behaviour of the perturbative series. The $r^{\lambda^\pm}{\rm e}^{-r}$ terms
in our transseries solutions therefore convert into $n^{\lambda^\pm-1} n!$ asymptotic contributions to the
$(n+1)$th term of the perturbative series. Terms with the two possible values of $\lambda$
must be combined to obtain real results and give in particular a slowly oscillating factor
$\cos({\rm Im}(\lambda^+)\ln(n) + \phi)$. The dependence of this phase on the logarithm of~$n$ means that
we would have to reach quite high orders to have an unambiguous sign of the associated oscillations. A~more
speculative point is that such singularities of the Borel transforms could be related to singularities of the
propagator at points proportional to the renormalisation group invariant mass~\cite{BeCl16}. The complex values
of $\lambda^\pm$ could imply complex values for the corresponding position of the singularities of the
propagator: such unusual analyticity properties of the propagator have been proposed as signs of confinement of
the corresponding field~\cite{HaKo2021,St02,St95}.

\section{Conclusion}

In this paper we have started the resurgent analysis of the Ward--Schwinger--Dyson equations for the $\phi^3_6$
model. This is a clear illustration of the power of this new approach to quantum field theory to address
asymptotic properties of the perturbative series. The exponents we compute are totally inaccessible to the simple
minded considerations of specific graphs that allowed to locate renormalon singularities of the Borel transform.
Classical field configurations which could reproduce such singularities in a
semiclassical expansion would have serious difficulties to explain such exponents.

It is clear that the situation is much more intricate than for the Wess--Zumino model studied
in~\cite{BeCl16}, where infinite families of possible transseries deformations could be readily obtained.
Already at the level we consider here of the nearest singularities of the Borel transform, we have a pair of complex conjugated exponents in the transseries
expansion. At the following levels, we would have three different objects of the type $F_2$, which could
mean that three different exponents are possible at level $[-2]$. And we cannot neglect
the possibility that higher order corrections to the Ward--Schwinger--Dyson equations have an influence on this
whole picture, since some aspects of our computations depend on the precise way we have done the infrared
rearrangements. It is therefore our hope that new constraints can be obtained that would allow us to tame this
proliferation of new series and use them to study
non-perturbative effects for this model through these methods.

It has recently been remarked that in a quite interesting special case, where the field is in a~bi-adjoint
representation, the first coefficient of the $\beta$ function vanishes, while the theory remains asymptotically
free due to the sign of the second coefficient~\cite{Gra2020}. In this case, the transseries solution would
include powers of $\exp\big( r^2 \big) = \exp\big( 1/a^2\big) $, meaning that singularities appear only for the Borel plane
dual to a variable $ u = r^2$ or some equivalent one. Such a study could reveal new aspects of resurgence
studies, but needs a knowledge of the $\beta$ function at higher loop for us to be able to control the precise
transmonomial appearing in the expansion of the renormalisation group function.

This is but a first step in the study of non-perturbative effects in this particular quantum
field theory, but it nevertheless presents results not accessible by other methods.

\subsection*{Acknowledgements}
The authors feel indebted to Dirk Kreimer for his many inspirational works and find befitting that this work be
published in a special issue in his honour.

\pdfbookmark[1]{References}{ref}
 \LastPageEnding

\end{document}